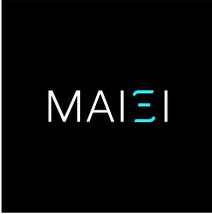

# Montreal AI Ethics Institute

*An international, non-profit research institute helping humanity define its place in a world increasingly driven and characterized by algorithms*

*__Website__: https://montrealethics.ai*
*__Newsletter__: https://aiethics.substack.com*

# The Gray Rhino of Pandemic Preparedness
*Proactive digital, data, and organizational infrastructure to help humanity build resilience in the face of pandemics*

## Prepared for:

Privacy & Pandemics: Responsible Uses of Technology and Health Data During Times of Crisis — An International Tech and Data Conference

October 27-28, 2020

Future of Privacy Forum, National Science Foundation,
Duke Sanford School of Public Policy, and Intel Corporation

## Prepared by:

**Abhishek Gupta**
Founder and Principal Researcher, Montreal AI Ethics Institute
Machine Learning Engineer and CSE Responsible AI Board Member, Microsoft
abhishek@montrealethics.ai
https://atg-abhishek.github.io

**Word Count:** 999 words (excluding references and section headings)

**Introduction:**

COVID-19 has exposed glaring holes in our existing digital[1], data[2], and organizational[3] practices. Researchers ensconced in epidemiological and human health work have repeatedly pointed out how urban encroachment, climate change, and other human-triggered activities and patterns are going to make zoonotic pandemics more frequent and commonplace[4]. The **Gray Rhino**[5] mindset provides a useful reframing (as opposed to viewing pandemics such as the current one as a *Black Swan*[6] event) that can help us recover faster from these (increasingly) frequent occurrences and build resiliency in our digital, data, and organizational infrastructure. Mitigating the social and economic impacts of pandemics can be eased through building infrastructure that elucidate leading indicators via passive intelligence gathering so that responses to containing the spread of pandemics are not blanket measures; instead, they can be fine-grained allowing for more efficient utilization of scarce resources and minimizing disruption to our way of life.

Yet, pervasive monitoring poses significant privacy[7] and trust challenges[8] limiting the efficacy of such measures. In addition, ad-hoc approaches further diminish trust in these measures. Participatory design[9] and proactive development to generate preparedness will not only elicit higher levels of trust, but also help to build infrastructures that are tailored better for the communities that they are meant to serve and help us better weather the harmful societal impacts of pandemics.

**Digital infrastructure interventions:**

From a digital infrastructure standpoint, continual investments in extending the reach of networks and internet access to the "last-mile" areas that are as of yet uncovered is important if we're to embark on helping **all** communities recover swiftly by sharing with them the necessary tools that allow them to function during periods of distancing and other protective measures mandates by health authorities. Sparsely connected regions have faced a disproportionate burden of this current pandemic[10] due to their inability to access digital services, hindering their ability to work remotely (to the extent that their workplaces allowed for that).

---

[1] Ramsetty, A., & Adams, C. (2020). Impact of the digital divide in the age of COVID-19. Journal of the American Medical Informatics Association, 27(7), 1147-1148.
[2] Holmdahl, I., & Buckee, C. (2020). Wrong but useful—what covid-19 epidemiologic models can and cannot tell us. New England Journal of Medicine.
[3] Horton, R. (2020). Offline: COVID-19 and the NHS—"a national scandal". Lancet (London, England), 395(10229), 1022.
[4] World Health Organization. (2003). Climate change and human health: risks and responses. World Health Organization.
[5] Wucker, M. (2016). The gray rhino: How to recognize and act on the obvious dangers we ignore. Macmillan.
[6] Mazzoleni, S., Turchetti, G., & Ambrosino, N. (2020). The COVID-19 outbreak: From "black swan" to global challenges and opportunities. Pulmonology, 26(3), 117.
[7] Rowe, F. (2020). Contact tracing apps and values dilemmas: A privacy paradox in a neo-liberal world. International Journal of Information Management, 102178.
[8] Woskie, L. R., & Fallah, M. P. (2019). Overcoming distrust to deliver universal health coverage: lessons from Ebola. bmj, 366, l5482.
[9] Gupta, A., & De Gasperis, T. (2020). Participatory Design to build better contact-and proximity-tracing apps. arXiv preprint arXiv:2006.00432.

Parallel investments in extending high-speed internet via existing wired and wireless networks can help reach sparsely populated regions[11]. This will enable better data collection, leading indicators generation, and tailoring policy responses to combat the pandemic in these regions. Reopening of the economy can also be done in an informed manner rather than the current trial-and-error approach that has led to recurring ebbs and flows of cases[12].

**Data infrastructure interventions:**

Here we approach the fundamental tension that has plagued the current technological deployment of solutions to combat COVID-19: privacy intrusions due to an expansion of surveillance infrastructure, whether benign[13] or one with ulterior motives[14]. There are three measures that can be adopted to mitigate these concerns: transparency in the design, development, and deployment of contact-tracing and other technologies, use of privacy-preserving technologies like differential privacy (DP)[15] as opposed to ad-hoc anonymization and other unproven methods, and proactive and premeditated data trusts run by communities themselves.

*Transparency*

Adopting transparent design, development, and deployment of digital contact-tracing has been shown to be better received than taking a closed-door approach as seen in the face-off between PEPP-PT[16] and DP-3T[17] protocols in Europe. The Apple-Google toolkit when adopted by governments[18] has also shown to be more effective in part (though potentially confounded by higher interoperability and ease of use) because of the technical standards[19] and considerations being made public ahead of time. When the design choices are made public ahead of time, iterated open under public scrutiny, and have a tracked version history (via a service like Github) as was used by DP-3T[20], it engenders more trust from users and increases the adoption of the technology bolstering its efficacy.

---

[10] Whitacre, B., & Gallardo, R. (2020). COVID-19 lockdowns expose the digital have-nots in rural areas – here's which policies can get them connected. Retrieved 8 September 2020, from https://theconversation.com/covid-19-lockdowns-expose-the-digital-have-nots-in-rural-areas-heres-which-policies-can-get-them-connected-144324

[11] Pereira, J. P. R. (2016). Broadband access and digital divide. In New advances in information systems and technologies (pp. 363-368). Springer, Cham.

[12] Xu, S., & Li, Y. (2020). Beware of the second wave of COVID-19. The Lancet, 395(10233), 1321-1322.

[13] Seshadri, D. R., Davies, E. V., Harlow, E. R., Hsu, J. J., Knighton, S. C., Walker, T. A., ... & Drummond, C. K. (2020). Wearable sensors for COVID-19: A call to action to harness our digital infrastructure for remote patient monitoring and virtual assessments. Frontiers in Digital Health, 2, 8.

[14] McGee, P., Murphy, H., & Bradshaw, T. (2020). Coronavirus apps: the risk of slipping into a surveillance state.

[15] Dwork, C., & Roth, A. (2014). The algorithmic foundations of differential privacy. Foundations and Trends in Theoretical Computer Science, 9(3-4), 211-407.

[16] Veale, M. (2020). Official PEPP-PT severance notice from ETH Zürich. Retrieved 8 September 2020, from https://twitter.com/mikarv/status/1251432072507465728?s=20

[17] Clarke, L. (2020). PEPP-PT vs DP-3T: The coronavirus contact tracing privacy debate kicks up another gear. Retrieved 8 September 2020, from https://tech.newstatesman.com/security/pepp-pt-vs-dp-3t-the-coronavirus-contact-tracing-privacy-debate-kicks-up-another-gear

*Privacy-preserving techniques*

Time[21] and again[22], it has been demonstrated that ad-hoc anonymization doesn't offer the most robust guarantees in terms of personal data protection. Yet, it is the dominant technique mentioned whenever discussions on privacy arise. Whether in popular media or in policy-makers' offices, a more thorough understanding of techniques like DP, which is able to provide mathematically verifiable bounds[23] on the protection of personal data, will be essential. More so, practical and **highly public** demonstrations using DP, for example the push from the US Census in 2020[24], will showcase that it is not a theoretical construct and can be scaled to a national level. Additionally, aligning language in privacy legislations and regulations that are closer to how such techniques might be realized in practice[25] will help to strengthen the case for adoption of these industry best practices.

*Data trusts*

Ad-hoc creation of data trusts tends to create discord and distrust in the communities that these solutions are meant to serve[26]. *A priori* creation of stewardship mechanisms for community-collected data is essential. Such an approach also creates a transparent *contract* between those who are tasked with managing the data and the data subjects. It also affords the opportunity for engaged discussion arriving at agreements that are aligned with the values upheld by the communities[27].

---

[18] Busvine, D., & Rinke, A. (2020). Germany flips to Apple-Google approach on smartphone contact tracing. Reuters, April, 26.
[19] Privacy-Preserving Contact Tracing - Apple and Google. (2020). Retrieved 12 September 2020, from https://covid19.apple.com/contacttracing
[20] Troncoso, C. (2020). DP-3T/documents. Retrieved 8 September 2020, from https://github.com/DP-3T/documents
[21] Narayanan, A., & Shmatikov, V. (2008, May). Robust de-anonymization of large sparse datasets. In 2008 IEEE Symposium on Security and Privacy (sp 2008) (pp. 111-125). IEEE.
[22] Ohm, P. (2010). What the Surprising Failure of Data Anonymization Means for Law and Policy.
[23] Tschantz, M. C., Kaynar, D., & Datta, A. (2011). Formal verification of differential privacy for interactive systems. Electronic Notes in Theoretical Computer Science, 276, 61-79.
[24] Bureau, U. (2020). Disclosure Avoidance and the 2020 Census. Retrieved 8 September 2020, from https://www.census.gov/about/policies/privacy/statistical_safeguards/disclosure-avoidance-2020-census.html
[25] Caron, M. S., & Gupta, A. (2020). Response to Office of the Privacy Commissioner of Canada Consultation Proposals pertaining to amendments to PIPEDA relative to Artificial Intelligence. arXiv preprint arXiv:2006.07025.
[26] Cohen, A., & Gupta, A. (2020). Report prepared by the Montreal AI Ethics Institute In Response to Mila's Proposal for a Contact Tracing App. arXiv preprint arXiv:2008.04530.

**Organizational infrastructure interventions:**

The creation of well-established playbooks and organizational measures such as dedicated bodies and committees that are composed of the necessary experts who have a long history of engagement on the technical, scientific, and social aspects of pandemic management helps to *bake-in* intelligence into the process through continual learning and iteration, something that is proving to be challenging as ad-hoc measures are springing up across the world[28]. Unaware of the work being done by others, and duplicating research[29] and creating best practices from scratch has led to uneven responses[30]. In a pandemic, such responses are particularly harmful since they limit the effectiveness of the measures undertaken by everyone else.

**Future Directions:**

Ultimately, embodying this *Gray Rhino* mindset has the benefit of ushering in an era of pandemic preparedness that taps into the best practices across digital, data, and organizational infrastructure that shifts the paradigm from being reactive to proactive. It is also centered on capacity building in a deliberate and continual manner to reduce uncertainty and the severity of negative consequences when a pandemic strikes. Respecting ethical considerations including privacy, data rights, context, and non-discrimination through the empowerment of people who are disproportionately affected by pandemics will help us utilize technology that truly does benefit everyone.

---

[27] Gupta, A., & De Gasperis, T. (2020). Participatory Design to build better contact-and proximity-tracing apps. arXiv preprint arXiv:2006.00432.
[28] Monitor, I. L. O. (2020). COVID-19 and the world of work.
[29] World Health Organization. (2020). Accelerating a safe and effective COVID-19 vaccine. Retrieved 14 September 2020, from https://www.who.int/emergencies/diseases/novel-coronavirus-2019/global-research-on-novel-coronavirus-2019-ncov/accelerating-a-safe-and-effective-covid-19-vaccine
[30] Monitor, I. L. O. (2020). COVID-19 and the world of work.